# Atomic scale electronic structure of the ferromagnetic semiconductor $Cr_2Ge_2Te_6$

Zhenqi Hao[1,‡], Haiwei Li[1,‡], Shunhong Zhang[2,‡], Xintong Li[1], Gaoting Lin[3,4], Xuan Luo[3], Yuping Sun[3,5,6], Zheng Liu[2*] and Yayu Wang[1,7*]

[1]*State Key Laboratory of Low Dimensional Quantum Physics, Department of Physics, Tsinghua University, Beijing 100084, China*
[2]*Institute for Advanced Study, Tsinghua University, Beijing 100084, China*
[3]*Key Laboratory of Materials Physics, Institute of Solid State Physics, Chinese Academy of Sciences, Hefei 230031, China*
[4]*University of Science and Technology of China, Hefei, 230026, China*
[5]*High Magnetic Field Laboratory, Chinese Academy of Sciences, Hefei 230031, China*
[6]*Collaborative Innovation Centre of Advanced Microstructures, Nanjing University, Nanjing 210093, China*
[7]*Collaborative Innovation Center of Quantum Matter, Beijing 100084, China*

[‡]These authors contributed equally to this work.

[*]Email: zheng-liu@tsinghua.edu.cn, yayuwang@tsinghua.edu.cn

ABSTRACT: $Cr_2Ge_2Te_6$ is an intrinsic ferromagnetic semiconductor with van der Waals type layered structure, thus represents a promising material for novel electronic and spintronic devices. Here we combine scanning tunneling microscopy and first-principles calculations to investigate the electronic structure of $Cr_2Ge_2Te_6$. Tunneling spectroscopy reveals a surprising large energy level shift and change of energy gap size across the ferromagnetic to paramagnetic phase transition, as well as a peculiar double-peak electronic state on the Cr-site defect. These features can be quantitatively explained by density functional theory calculations, which uncover a close relationship between the electronic structure and magnetic order. These findings shed important new lights on the microscopic electronic structure and origin of magnetic order in $Cr_2Ge_2Te_6$.

1. **Introduction**

Ferromagnetic semiconductors (FMSs) are of great technological importance due to their potentials in utilizing the spin degree of freedom to realize new and better functions. A variety of FMS-based applications have been proposed and realized in information storage, computation, and communication [1-3]. One promising approach to realize FMS is by diluted magnetic doping into well-known semiconductors, as exemplified by Mn doped GaAs [4,5]. Although tremendous progresses have been made along this direction, most diluted magnetic semiconductors encounter the complications from disorders, inhomogeneity, and unintentional charge doping [6]. Therefore, intrinsic ferromagnetism in pure phase semiconductors becomes a more attractive choice for FMS-based devices. However, due to the lack of itinerant electrons to mediate ferromagnetic (FM) order, most intrinsic semiconductors are either paramagnetic (PM) or exhibit anti-ferromagnetic (AFM) order resulting from the super-exchange mechanism.

In recent years, the $Cr_2X_2Te_6$ (X = Si, Ge, Sn) series compounds have attracted much attention because some of them are intrinsic FMSs with Curie temperature ($T_C$) as high as 70 K [7-17]. Moreover, their van der Waals (vdW) type layered structure facilitates the fabrication of nanoscale devices and integration with various layered transition metal dichalcogenides with novel electronic [18,19], magnetic [20], optical/optoelectric [21-24], piezoelectric [25], magnetoresistive [26], and superconducting [27] properties. More interestingly, it has been reported recently that FM order exists in atomic layers of pristine $Cr_2Ge_2Te_6$, and the $T_C$ can be tuned easily by an applied magnetic field [28]. This surprising discovery indicates that $Cr_2Ge_2Te_6$ is a unique model system for studying two-dimensional magnetic phenomena and novel spintronic applications. However, so far very little is known about the microscopic properties of $Cr_2Ge_2Te_6$, such as the size of the energy gap ($E_g$), the energy and spatial distribution of electron density of states (DOS), and the characteristic impurity electronic states. More importantly, the origin of the FM order in $Cr_2Ge_2Te_6$ and its relationship to the microscopic electronic structure are still elusive. Owing to its ability of simultaneously mapping out the local structural and electronic properties, scanning tunneling microscopy (STM) serves as a perfect experimental technique to answer these questions.

In this work, we use STM to investigate the structural and electronic properties of crystalline $Cr_2Ge_2Te_6$ down to the atomic scale. The energy gap around the Fermi level ($E_F$) shows an unexpected decrease as the sample is cooled to below $T_C$, which is accompanied by a pronounced upwards shift of the unoccupied electronic levels. The clover-shaped defect at the Cr site exhibits two distinct spectral peaks within the semiconducting gap. These results can be quantitatively reproduced by density functional theory (DFT) calculations, which reveal a close relationship between the microscopic electronic structure and magnetic order in $Cr_2Ge_2Te_6$.

## 2. Experimental

The $Cr_2Ge_2Te_6$ single crystals used here were grown by the self-flux technique, as described in a previous work [7]. The sample has the same magnetic property as that reported in Ref. [7], with a $T_C \sim 68$ K. The STM experiments were performed in ultra-high vacuum (UHV) environment with pressure down to $10^{-10}$ mbar. The $Cr_2Ge_2Te_6$ single crystal is cleaved at 77 K in the UHV preparation chamber, and then transferred to the STM chamber for measurement. An electrochemically etched Tungsten tip is prepared and calibrated on an atomically clean Au(111) surface before being used on the sample [29]. The STM topography is obtained in the constant current mode, and the differential conductance (d$I$/d$V$) measurements are performed using an AC lock-in method with modulation frequency 423 Hz.

## 3. Results and discussion

Figure 1a and 1b display the side-view and top-view of the schematic crystal structure of $Cr_2Ge_2Te_6$ respectively. Each crystalline layer consists of edge-sharing $CrTe_6$ octahedrons forming a honeycomb lattice and Ge-dimers residing at the central hexagon. The weak vdW bonding between neighboring layers allows an easy cleavage of the crystal along the gray plane illustrated in Fig. 1a, which exposes a charge neutral surface of the Te layer. Figure 1c shows a large scale topographic image obtained on the surface of cleaved $Cr_2Ge_2Te_6$ with sample bias voltage $V = -2.5$ V. The overall pattern is a well-ordered honeycomb lattice with four defects in the form clover-shaped bright spots. Figure 1d displays a high resolution image on a defect free area taken with a decorated tip, which helps enhance the atomic contrast. By comparing with the schematic structure in Fig. 1b, it can be clearly seen that each bright spot corresponds to a surface Te atom, and the Cr atoms reside at the center of three nearest-neighboring bright spots. The Ge-dimer sites are dark in the topograph, leading to the hollows of the honeycomb lattice. Figure 1e is the theoretically simulated surface topography based on *ab initio* calculations, which reproduces the experimentally observed image very well. The calculation details will be discussed later.

To probe the electronic structure of $Cr_2Ge_2Te_6$, we perform the $dI/dV$ spectroscopy, which reflects the local electron DOS of the sample. On an area far from the defect, the $dI/dV$ spectra are highly uniform in space with no discernable variations at different lattice sites. The upper panel of Fig. 2a shows the spatially-averaged $dI/dV$ spectrum on a defect-free area measured at $T = 50$ K $< T_C$, when the sample is in the FM state, whereas the lower panel shows that measured at $T = 77$ K $> T_C$ in the PM state. The main features of the two curves are qualitatively the same. There is a large energy gap around the $E_F$, and there are multiple DOS peaks outside the gap. However, a closer examination of the two spectra reveals something highly unusual. When the sample is cooled from the PM state to FM state, the overall lineshape remains the same but the characteristic peaks on the positive bias side (corresponding to the unoccupied state) move to higher energy, as indicated by the broken lines. Another totally unexpected feature is that the energy gap in the FM state is actually smaller than that of the PM state, as shown clearly by the logarithmic plot in Fig. 2b. The blue dashed lines are linear fits to the leading edge of the d$I$/d$V$ curves, which is commonly used to define $E_g$ of a semiconductor. The gap size measured at $T = 50$ K is around 0.74 eV, whereas that measured at $T = 77$ K is around 0.86 eV. The decrease of $E_g$ across the PM to FM transition is by itself unusual, and it is even more surprising given the upwards shift of the unoccupied DOS peaks in the FM state.

To understand the observed STM results on $Cr_2Ge_2Te_6$, we have performed first-principles calculations within the DFT implemented in the VASP code [30]. For the lattice structure, we find that by combining the PBE functional [31], the D2 empirical correction for vdW interaction [32], and a FM spin-density configuration, the converged numerical results ($a = b = 6.89$ Å, $c = 20.21$ Å) agree well with the experimentally determined bulk lattice constants ($a = b = 6.83$ Å, $c = 20.56$ Å). For the electronic structure, the vdW-type interlayer coupling does not play an important role, and the calculated real-space electron density contour above the surface of a monolayer well reproduces the STM topography (Fig. 1e). The DOS of the monolayer at the PBE+FM level already characterizes the FMS ground state, despite a relatively small $E_g \sim 0.40$ eV. The underestimated gap size can be improved by introducing a +$U$ correction [33] ($U_{eff} = 1.5$ eV, $E_g \sim 0.45$ eV, see total DOS in top panel of Fig. 3b). We note that hybrid Hartree-Fock/DFT calculation [34,35] can produce a gap size ($E_g \sim 0.70$ eV) even closer to the experimental value. However, a comprehensive hybrid functional study is computationally expensive, so we leave itt for future investigations. The local octahedral $CrTe_6$ coordination leads to a primary crystal-field splitting ($\Delta_{oct}$) between the Cr $t_{2g}$- and $e_g$-orbitals (Fig. 3d). By projecting the DFT+$U$ eigenstates onto the atomic orbitals along the local octahedral axes, we can clearly identify DOS peaks dominated by Cr $t_{2g}$- and $e_g$-orbitals around the band edge (Fig. 3b). The energy difference between the occupied and empty $t_{2g}$-orbitals represents the strength of onsite Coulomb repulsion. The energy difference between the two empty $e_g$ sets with opposite spin polarization represents the strengths of Hund's rule coupling ($J_H$). The combination of crystal-field splitting and Coulomb interaction dictates a valence band edge dominated by the spin polarized $t_{2g}$-orbitals, and a conduction band edge dominated by $e_g$-orbitals with the same spin polarization (Fig. 3b). A small fraction of DOS from Te $p$-orbitals can also be observed around the band edge as a result of $p$-$d$ hybridization, which in turn mediates the magnetic coupling between two nearest-neighbor $Cr^{3+}$ ions. The FM ground state can be empirically understood by the Goodenough-Kanamori rules [36,37] - indeed, the Cr-Te-Cr bond angle is close to 90 degree. One can also rationalize the stability of FM order by taking into account the Cr $d$-orbitals only. The key point is that under the bonding geometry of $Cr_2Ge_2Te_6$, the $e_g$ and $t_{2g}$ orbitals of the neighboring Cr atoms are nonorthogonal. As shown in the lower panel of Fig. 3d, if we imagine a virtual electron hopping from the $t_{2g}$-orbital of one Cr to the $e_g$-orbital of neighboring Cr, the excitation energy of this virtual process is lower when the two Cr-sites have the parallel spin polarization due to the Hund's rule coupling, leading to FM superexchange.

Regarding the change of $dI/dV$ spectrum across $T_C$, it is well-known that in conventional semiconductors $E_g$ typically *decreases* with increasing temperature at a rate ~ 0.1 meV/K [38], i.e. $\Delta E_g/k_B\Delta T$ ~ -1. Such change is a natural consequence of the temperature-dependent lattice dilatation and electron–lattice interaction. However, here $E_g$ *decreases* by over 100 meV when the temperature increases by 27 K, which gives $\Delta E_g/k_B\Delta T$ ~ 40. Our atomically resolved STM topography does not show visible structural change across the FM-to-PM phase transition, so we can exclude the influence of lattice thermodynamics. We neither expect the interaction strength to change by 100 meV within 27 K. Previous optical studies on Mott insulators, such as LaTiO$_3$ (Ref. [39]), YTiO$_3$ (Ref. [40]) and Yb$_2$V$_2$O$_7$ (Ref. [41]), showed that their electronic structure has little temperature dependence.

We propose that the most likely origin of the anomalous gap variation is magnetically driven. A closely related example is the so-called "Slater insulator", in which the formation of AFM order open a band gap. In fact, Cr$_2$Ge$_2$Te$_6$ realizes Slater's idea in a complimentary way – the formation of FM order renormalizes the electronic structure and reduces the band gap. The schematics of Fig. 3d provide an intuitive explanation for the gap variation. However, it does not take into account the band formation process, thus cannot explain the observed upwards shift of conduction band peaks upon the formation of FM order. To quantitatively evaluate the relation between magnetic order and electronic structure, we designed a series of DFT+$U$ calculation by purposely perturbing the FM order and tracing the DOS evolution. Figure 3a shows a monolayer supercell containing eight Cr sites. We start from the FM configuration, and then artificially flip the spin of one site, two sites and three sites. With four sites flipped, we effectively simulate an AFM order. It is clear that as the magnetic configuration deviates from the FM order, the e$_g$ peak has a visible downshift towards $E_F$ (Fig. 3b). On the other hand, the total width of the e$_g$ band shrinks, which reflects that conduction electrons are more itinerant in a FM background, whereas an AFM background tends to block the motion. A zoom-in plot around the band edge (Fig. 3c) reveals that $E_g$ is smallest for the FM configuration and increases progressively when the spin configuration deviates from the FM order. The gap size difference between the FM and AFM configurations is around 100 meV. In the realistic PM state slightly above $T_C$, Cr spins are fluctuating with short-range FM correlation and the electronic structure can be interpreted as a result of the combination of different magnetic orders. Based on the calculation results, we conclude that these dynamic deviations from a perfect FM order frustrate electron motion, shrink the bandwidth and consequently lead to a larger gap.

Now that the electronic structure of pristine $Cr_2Ge_2Te_6$ is understood, we will focus on the impurity electronic state on the defect. In semiconductor physics, defects have important implications for the physical properties and device performances. Figure 4a shows the zoomed-in image on the clover-shaped defect, and the center of the defect can be determined to be located at the Cr site. The blue curve in Fig. 4c represents the local DOS taken at the defect center at $T = 50$ K. Compared to that on the defect-free area (black curve), the most prominent features are two sharp in-gap state peaks located at bias voltage -0.2 V and 0.5 V respectively. The spectrum outside the energy gap is nearly identical to that of the pristine phase. These features can be well reproduced in the DFT calculation by considering Ge substitution of a Cr atom ($Ge_{Cr}$). Figure 4b shows the simulated STM topography around the $Ge_{Cr}$ defect, and Fig. 4e displays the calculated DOS in a supercell containing one $Ge_{Cr}$ defect. The nature of the two in-gap states can be better seen from the band structure. In Fig. 4d, we observe two nearly flat bands within the gap, which can be unambiguously attributed to the defect-induced in-gap states. More interestingly, they carry opposite spins, resembling a Zeeman splitting due to proximity to the nearby Cr spins, which have short-range FM correlation even in the PM state. Therefore, the $Ge_{Cr}$ impurity site holds net spin opposite to the FM background formed by Cr, which does not exist in nonmagnetic semiconductors. One consequence is that via spin splitting, the $Ge_{Cr}$ defect is self-compensated, i.e. it provides one donor level and one acceptor level simultaneously. This defect thus does not affect the insulating performance of $Cr_2Ge_2Te_6$ at low temperature, which can be important in device applications. The defect also represents an ideal two-level system embedded in a FMS that may be used in information storage and computation.

## 4. Conclusion

In summary, we have performed atomic scale STM experiments on the cleaved crystal of $Cr_2Ge_2Te_6$, an intrinsic FMS with vdW-type layered structure. We found an interesting correlation between the electronic structure and magnetic order, which leads to a surprisingly large energy level shift and change of band gap across the FM to PM phase transition. These results provide important new insights into the microscopic electronic structure and magnetic order of $Cr_2Ge_2Te_6$. Theoretically, these new results reflect from an interesting angle the common wisdom – AFM has an insulating tendency, whereas FM has a metallization tendency. Similar gap reduction upon formation of FM order is expected to occur in other FM semiconductors as well.

ACKNOWLEDGMENT

This work was supported by the Basic Science Center Project of NSFC under grant No. 51788104, the MOST of China grant No. 2015CB921000. S.H.Z. and Z.L. acknowledge support from Tsinghua University Initiative Scientific Research Program and NSFC under Grant No. 11774196. S.H.Z. is supported by the National Postdoctoral Program for Innovative Talents of China (BX201600091) and the China Postdoctoral Science Foundation (2017M610858). G.T.L, X.L and Y.P.S thank the support of the National Key Research and Development Program grant No. 2016YFA0300404 and NSFC grant No. 11674326 and the Joint Funds of NSFC and the Chinese Academy of Sciences' Large-Scale Scientific Facility under contract U1432139. This work is supported in part by the Beijing Advanced Innovation Center for Future Chip (ICFC).

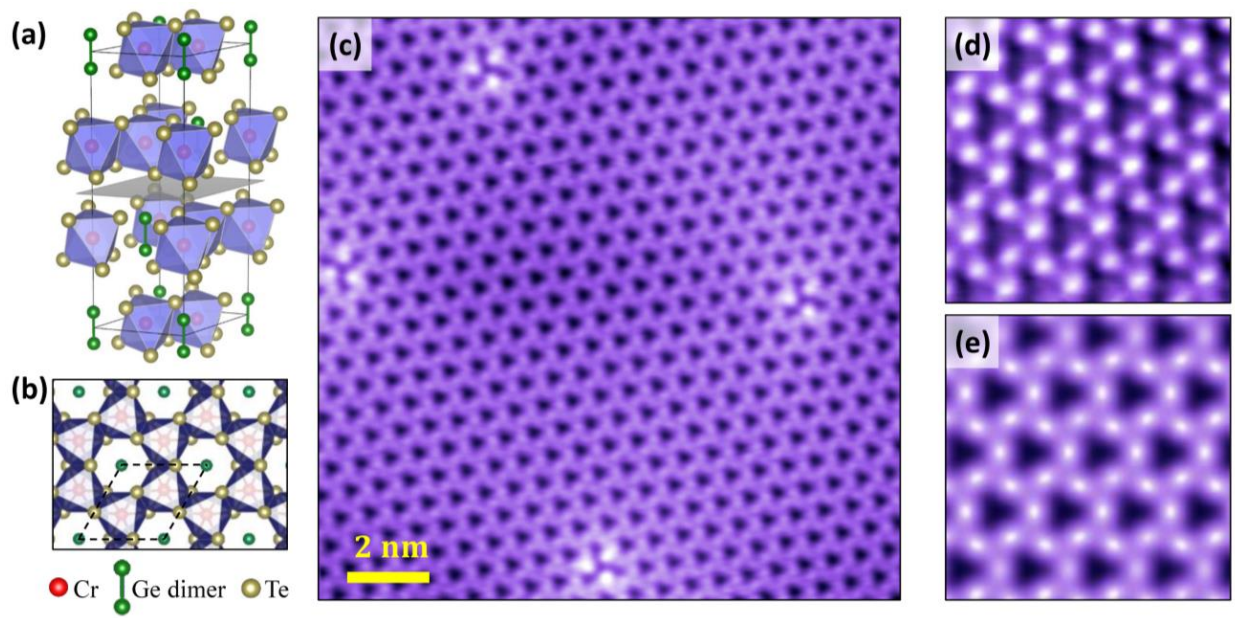

**Figure 1.** (Color online) Crystal structure and STM topography. (a) Side view of the $Cr_2Ge_2Te_6$ crystal structure. (b) Top view of the $Cr_2Ge_2Te_6$ crystal structure along the c-axis perpendicular to the plane. (c) A large scale STM topography of the cleaved $Cr_2Ge_2Te_6$ surface with four clover-shaped defects. (d) A zoomed-in image with the contrast enhanced by a decorated STM tip. (e) Simulated topography of a cleaved $Cr_2Ge_2Te_6$ surface based on the DFT calculations.

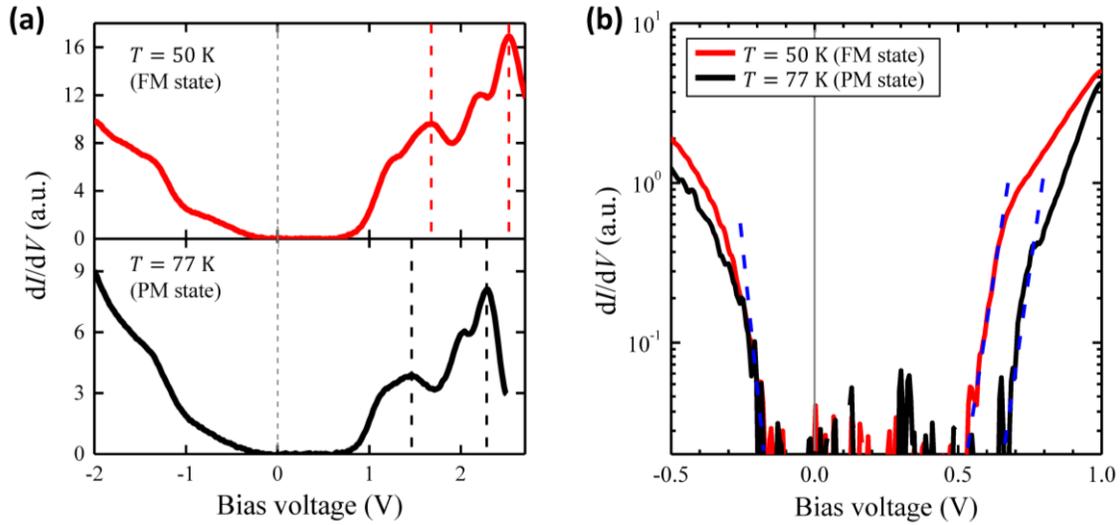

**Figure 2.** (Color online) Comparison of tunneling spectroscopy at ferromagnetic and paramagnetic states. (a) The tunneling spectroscopy on a defect free surface of $Cr_2Ge_2Te_6$ surface taken at $T =$ 50 K (upper panel) and 77 K (lower panel). The solid vertical line at zero bias indicate the position of $E_F$, and the broken lines indicate the upwards shift of unoccupied energy levels in the FM state. (b) The size of the energy gap is determined to be 0.74 eV and 0.86 eV for the FM state and PM state respectively.

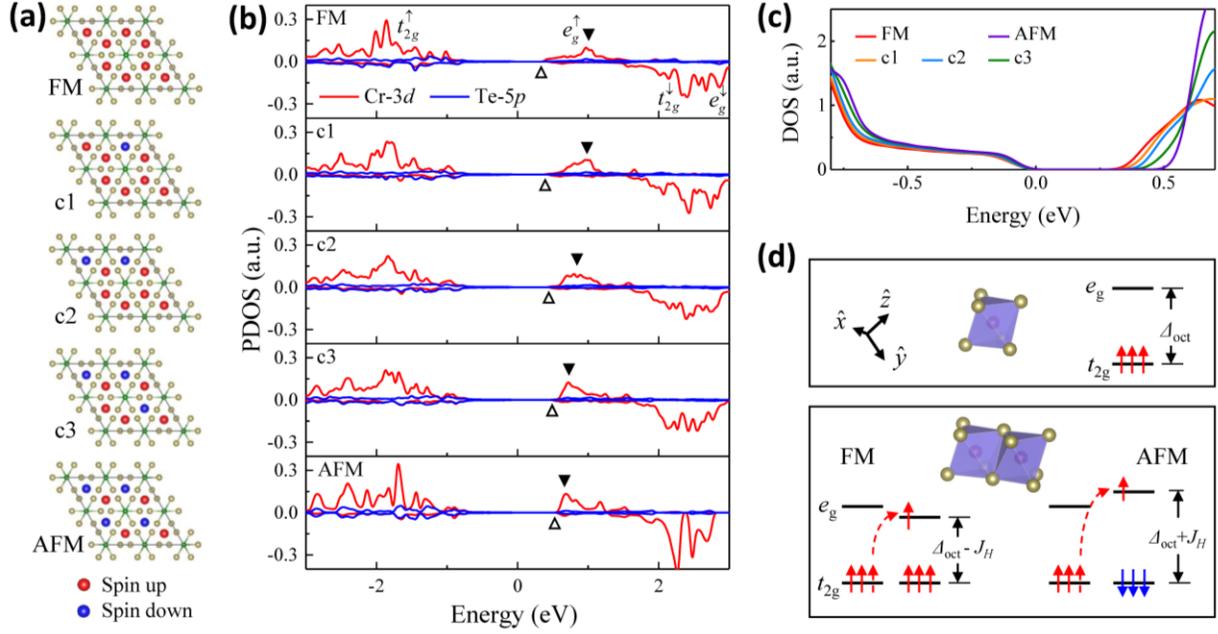

**Figure 3.** (Color online) First-principles calculation of different magnetic orders. (a) Five spin configurations used in the simulations. (b) Projected DFT+$U$ ($U$ = 1.5 eV) DOS of a monolayer $Cr_2Ge_2Te_6$ with different spin configurations. Positive and negative DOS represent the spin up and down channels respectively. The up and down triangles mark the conduction band edge and the main peak of the lower $e_g$ band, respectively. (c) Zoom-in DFT+$U$ total DOS for different spin orders around the gap region. (d) Upper panel: schematic illustration of the octahedral crystal field of Cr and the corresponding orbital splitting ($\Delta_{oct}$). Lower panel: the virtual electron hopping (illustrated by the dashed arrow) between the nearest $Cr^{3+}$ sites in the FM and AFM configurations.

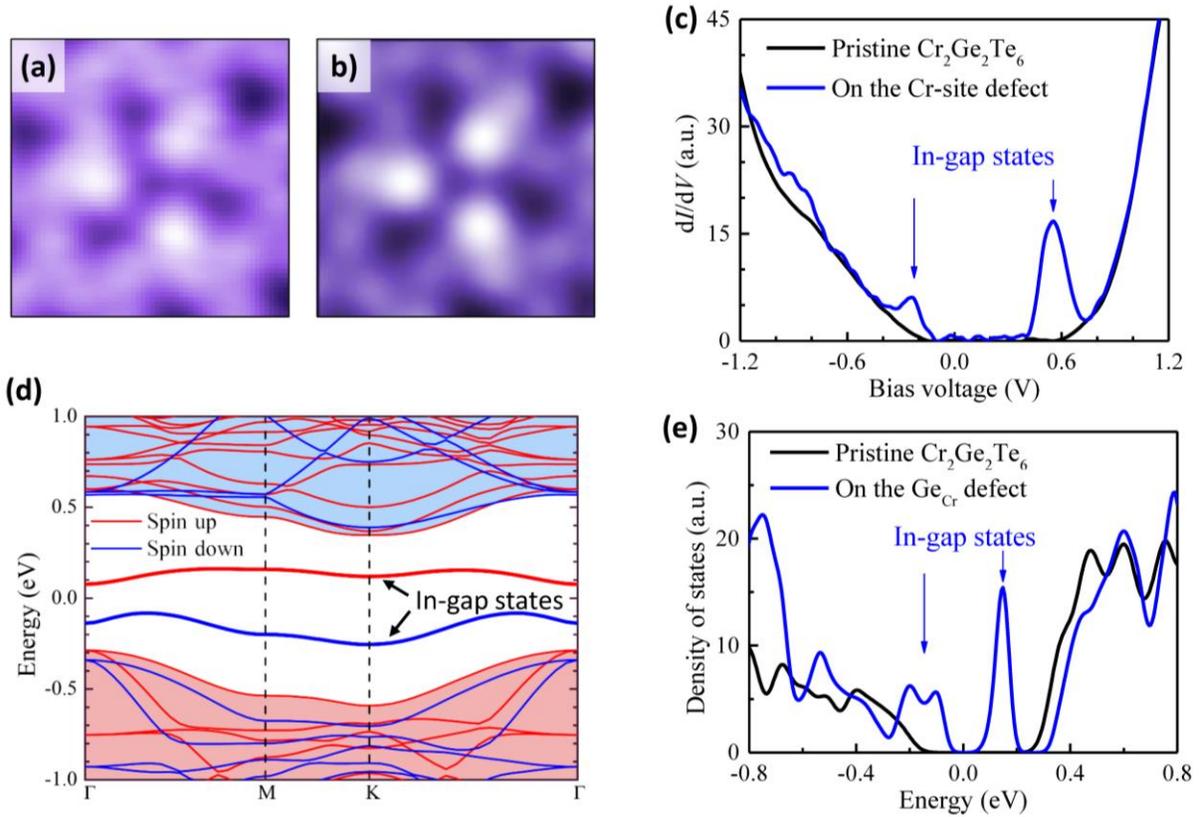

**Figure 4.** (Color online) Analysis of the defect on Cr site. (a) High-resolution STM topography on a Cr-site defect. (b) Simulated STM topography of the Ge$_{Cr}$ substitution defect based on DFT-calculated electronic state. (c) The tunneling spectra measured on the Cr-site defect (blue) and pristine Cr$_2$Ge$_2$Te$_6$ (black), which clearly reveal two peaks within the energy gap on the defect. (d) The calculated electronic band structure containing one Ge$_{Cr}$ defect in a 2×2 supercell. (e) The calculated total DOS on the defect and pristine phase.